\documentclass[twocolumn,showpacs,preprintnumbers,amsmath,amssymb]{revtex4}
\usepackage{graphicx}
\usepackage{dcolumn}
\usepackage{bm}

\begin{document}

\title{Effect of Soft Modes on the Shear Viscosity of Quark Matter}

\author{Takahiko Fukutome} \email{hiko_fukutome@world.odn.ne.jp}
\author{Masaharu Iwasaki} \email{miwasaki@cc.kochi-u.ac.jp}
\affiliation{Department of Physics, Kochi University, Kochi 780-8520, Japan}

\date{\today}

\begin{abstract}
We calculate shear viscosity of the quark matter at finite temperature and density. If we assume that the quark interacts with the soft mode, which is a collective mode of quark-antiquark pair, the self energy of the quark is calculated by using the quasi-particle random phase approximation. It is shown that its imaginary part is large and the mean free path of the quark is short. With the use of the Kubo formula, the shear viscosity of quark matter becomes small. The Reynolds number of the quark matter is estimated to be about $3\sim 30$. As the temperature increases, the shear viscosity increases gradually for $T>200{\rm MeV}$. Moreover it is shown that the shear viscosity is not sensitive to the chemical potential.
\end{abstract}


\maketitle

\section{\label{sec:intro}Introduction}

 The existence of quark matter or quark gluon plasma (QGP) has been predicted by Quantum Chromodynamics (QCD) which is the fundamental theory of hadronic world. It is supposed that the quark matter may exist in the early stage of the universe or in the core of the compact stars. Recently such a new state of matter has been explored in the high-energy heavy ion collisions (RHIC) \cite{IA05}-\cite{KA05}. Now it is very important to study the various physical properties of the quark matter in relation to these researches.

Several years ago one of the present authors (M.I.) studied the specific heat and compressibility of the quark matter by using the Nambu-Jona-Lasinio (NJL) model \cite{I04}. These quantities are enhanced near the critical temperature. On the other hand the transport coefficients, shear viscosity and heat conductivity, are also important in relation to the high-energy heavy ion collisions. Recently it was discovered that the quark matter seems to be almost perfect fluid above the critical temperature. Therefore the calculation of the shear viscosity has been paid attention by many authors from the various viewpoints \cite{HK85D}-\cite{GA06}, \cite{KW87}-\cite{NS05}(lattice QCD), \cite{KSS05}-\cite{HD05}(Black hole physics).

In this paper we try to calculate the shear viscosity of the quark matter by using the Kubo formula and the NJL model. We start to discuss the shear viscosity in terms of the Kubo formula \cite{K57}-\cite{Z74}, which is expressed by the correlation function of the energy-momentum tensor. Suppose that the quark system in the QGP is a free gas. Then the shear viscosity becomes {\it infinite} because of the strong correlation between their current densities. On the other hand the RHIC data suggests very small viscosity. In order to clarify the physical origin of this small viscosity, we consider the shear viscosity of quark sector with the use of the NJL model, which does not include gluons explicitly. In the previous paper (referred to as I), the quasi-particle random phase approximation (RPA) was applied to this formula and it was shown that the shear viscosity is represented by the quadratic form of the quark spectral function \cite{IOF07}. Consequently it is the main purpose of this paper to calculate the spectral function within the NJL model and obtain the values of the shear viscosity under the various temperature and density.

The spectral function is related to the self energy of the quark. In particular it depends on its imaginary part (or mean free path) strongly. What interaction does enhance the self energy? Here we are reminded of the soft mode, which was proposed as a precursor of the chiral phase transition by Hatsuda and Kunihiro long time ago \cite{HK85}. These modes grow strongly just above the critical temperature which is just the region we are interested in. We think that the collective mode couples strongly to the quark. We calculate the self energy of the quark due to the coupling to the soft modes. As is pointed out in \cite{HK94}, the spectral function of the soft mode decreases as the temperature increases. This means that the shear viscosity seems to be enhanced as the temperature increases. We will investigate the temperature dependence of the shear viscosity as well as the density dependence. Finally we calculate the Reynolds number of the quark matter in order to discuss the magnitude of the calculated shear viscosity.

In the next section, it is reviewed that the shear viscosity is expressed in terms of the quark spectral function, which is the main result of I. Then the spectral function is represented by the self energy of the quark (\S.3). In \S.4, the main part of this paper, we calculate the self energy by using the coupling to the soft modes in the quark matter. The numerical calculations are carried out and their results are discussed in the final section (\S.5).

\section{\label{sec:viscosity}Shear Viscosity in the NJL Model}

The Kubo formulas for the hydrodynamic transport coefficients are expressed in terms of the correlation function of the energy-momentum tensor $J_{xy}({\bf r},t)$. According to I, the formula for shear viscosity is given by
\begin{equation}
\eta\equiv \eta(\omega=0)=\left.-\frac{\mathrm{d}}{\mathrm{d}\omega}\text{\rm Im}\Pi^{\text{R}}(\omega)\right|_{\omega=+0}\;
\label{eq:def_viscosity},
\end{equation}%
where $\Pi^{\text{R}} (\omega)$ is a retarded Green's function obtained by analytic continuation of the following (imaginary time) Green's function: $\Pi^{\text{R}} (\omega) =\left.\Pi(i\omega_{n})\right|_{i\omega_{n}=\omega+i\delta}$,
\begin{equation}
\Pi(i\omega_n) =-\int_{0}^{\beta}\mathrm{d}\tau\mathrm{e}^{{-i\omega_{n}\tau}}\int\mathrm{d}{\bf r}\langle T_\tau(J_{xy}(\mathbf{r},\tau) J_{xy}(0,0))\rangle\;
\label{eq:im_Green}.
\end{equation}%
Here $\omega_{n}=2\pi nT\: (n=0,\pm 1,\pm 2, \cdots)$ is the Matsubara
frequency for bosons and $T_{\tau}$ means the (imaginary) time ordering operator. 

We take the NJL model for the quark matter in this paper \cite{HK94}. Then the Lagrangian density is given by
\begin{equation}
\mathcal{L}=\bar\psi(i\gamma \cdot \partial-m)\psi +g[(\bar\psi\psi)^2+(\bar\psi i\gamma_{5}\bm{\tau}\psi)^2]\;
\label{eq:NJL_Lag},
\end{equation}
where $\psi$ is the field operator for quarks and $g$ is their coupling constant. We assume that the current quark mass is zero ($m=0$) for convenience. The canonical energy-momentum tensor is read as
\begin{equation}
J_{xy} =\frac{i}{2}[\bar\psi\gamma^2\partial^1\psi-\partial^1\bar\psi\gamma^2\psi]\;
\label{eq:energy_mom}.
\end{equation}
When this equation is substituted into Eq.(\ref{eq:im_Green}), we have an infinite series for the correlation function $\Pi$. 

\begin{figure}
 \includegraphics[width=\linewidth]{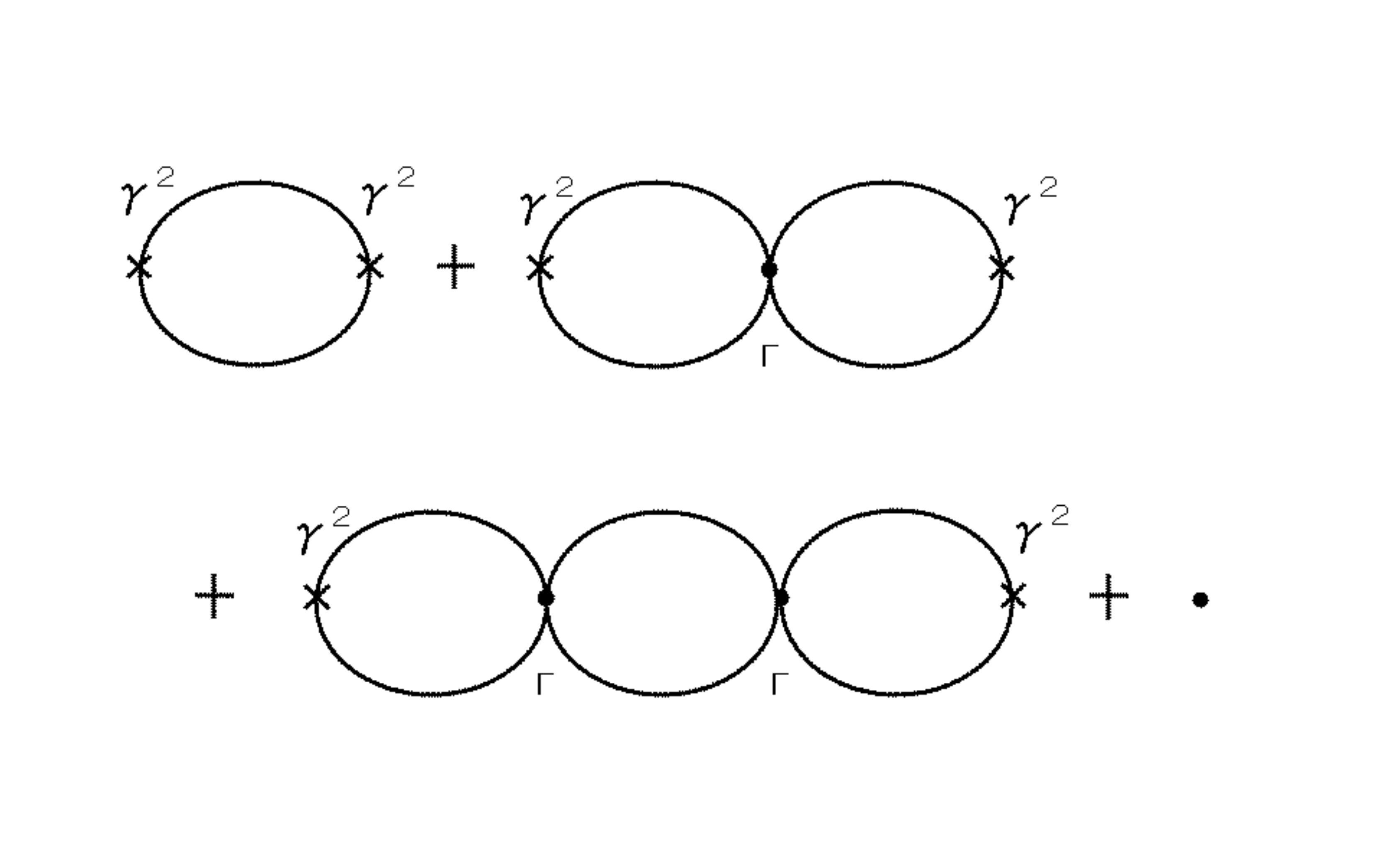}
  \caption{The ring diagrams for the correlation function $\Pi$ with
 $\Gamma=1$ or $i\gamma_5 \bm{\tau}$.}
 \label{fig:ring_diagrams_1}
\end{figure}

According to I, we approximate the higher orders by the ring diagrams shown in Fig.\ref{fig:ring_diagrams_1} (random phase approximation).
As is shown in I, the higher order terms which consists of two loops or more vanish. Because the leftist (rightist) loop of each diagram contains odd number of $\gamma^\mu(\mu=0,1,2,3)$ matrices and its trace vanishes. Thus only the first term of Fig.1 survives. The correlation function is written as 
\begin{eqnarray}
\Pi(i\omega_n)&=&\frac{1}{\beta}\sum_l\int\frac{\mathrm{d}^{3}p}{(2\pi)^3}{p_x}^2{\rm Tr}[{\gamma^2}G({\bf p},i\varepsilon_l+i\omega_n) \nonumber \\
& &{\gamma^2}G({\bf p},i\varepsilon_l)]\;
 \label{eq:Matsubara_sum},
\end{eqnarray}%
where $G({\bf p},i\varepsilon_l)$ denotes the dressed quark propagator. If the free propagator is substituted instead of the dressed one, we have an infinite viscosity. So finite viscosity comes from the finite life time of the quark or finite mean free path in the present model. The spectral representation for the dressed propagator is written as
\begin{equation}
G_{\alpha\beta}({\bf p},i\varepsilon_l) =\int_{-\infty}^\infty
 \frac{\mathrm{d}\varepsilon}{2\pi} \frac{\rho_{\alpha\beta}({\bf
 p},\varepsilon)}{i\varepsilon_l-\varepsilon}\;
 \label{eq:spectral_ft},
\end{equation}
where $\varepsilon_{l}=2\pi (n+1)T\: (n=0,\pm 1,\pm 2, \cdots)$ is the Matsubara frequency for fermions. Substituting this expression into the correlation function Eq.(\ref{eq:Matsubara_sum}), we obtain the expression of $\eta$ in terms of quark spectral function as follows:
\begin{equation}
\eta = -\frac{1}{2}\int_{-\infty}^{\infty}
 \frac{\mathrm{d}\varepsilon}{2\pi}
 \int\frac{\mathrm{d}^{3}p}{(2\pi)^3}{p_x}^2\frac{\partial n_{F}(\varepsilon)}{\partial
 \varepsilon}{\rm Tr}[\gamma^2\rho({\bf p},\varepsilon)\gamma^2\rho({\bf
 p},\varepsilon)]\;
 \label{eq:viscosity_spectral}.
\end{equation}
Here $n_{\text{F}}(\varepsilon)\equiv(1+\mathrm{e}^{\beta\varepsilon})^{-1}$ denotes the Fermi distribution function ($\beta=1/T$). The detailed derivation of this equation was given in I. The factor $\partial n_{\text{F}}/{\partial \varepsilon}$ in the right-hand side implies that the shear viscosity is crucially affected by the distribution function near the Fermi surface.

Next we calculate the spectral function of quarks. The spectral function is defined by
\begin{equation}
\rho({\bf p},\varepsilon)=i[G^{\text{R}}({\bf p},\varepsilon)-G^{\text{A}}({\bf p},\varepsilon)]\;,
\end{equation}
where the retarded and advanced Green's function are written as
\begin{subequations}
\begin{eqnarray}
 G^{\text{R}}({\bf p})&= G({\bf p},i\varepsilon_l=\varepsilon+i\delta)\;,
 \\
 G^{\text{A}}({\bf p})&= G({\bf p},i\varepsilon_l=\varepsilon-i\delta)\;.
\end{eqnarray}
\end{subequations}
The full Green's function $G$ is expressed in terms of the self energy $\Sigma$ as follows
\begin{equation}
G(p)^{-1}=G^{(0)}(p)^{-1}-\Sigma(p)\;
 \label{eq:GF_SE},
\end{equation}
where $G^{(0)}(p)$ denotes the free Green's function. Here we introduce the projection operators for positive and negative energy states, $\Lambda_{\pm}\equiv\frac{1}{2}(1\mp\gamma^{0}{\bm{\gamma}}\cdot\hat{{\bf p}})$. According to H.A.Weldon \cite{W00}, the free Green's function $G^{(0)}$ is written as a sum of two (positive and negative) parts.
\begin{equation}
 G^{(0)}(p)^{-1} =
 (p_+^0+p)\gamma^{0}\Lambda_{+}+(p_+^0-p)\gamma^{0}\Lambda_-\;
 \label{eq:Free_GF}\;,
\end{equation}
where $p_+^{0} \equiv i\varepsilon_l+\mu$ ($\mu=$chemical potential). Similarly, the self energy can be written as
\begin{subequations}
\label{eq:SE}
 \begin{eqnarray}
 \Sigma^{\text{R}} (p) &=& \Sigma_+ (p)(\gamma^0\Lambda_+)+\Sigma_-
 (p)(\gamma^0\Lambda_-)\;,
 \\
 \Sigma^{\text{A}} (p) &=& \Sigma_+^* (p) (\gamma^0\Lambda_+)+\Sigma_-^*
 (p)(\gamma^0\Lambda_-)\;.
 \end{eqnarray}
\end{subequations}
By substituting Eq.(\ref{eq:Free_GF}) and (\ref{eq:SE}) into Eq.(\ref{eq:GF_SE}), we obtain the expressions,
\begin{subequations}
\label{eq:GF_R&A}
\begin{align}
 G^{\text{R}} (p) &=& \frac{\gamma^0 \Lambda_-}{p_+^0 +p-\Sigma_+
 +i\delta}+\frac{\gamma^0 \Lambda_+}{p_+^0 -p-\Sigma_- +i\delta}\;,
 \\
 G^{\text{A}} (p) &=& \frac{\gamma^0 \Lambda_-}{p_+^0 +p-\Sigma_+^*
 -i\delta}+\frac{\gamma^0 \Lambda_+}{p_+^0 -p-\Sigma_-^* -i\delta}\;.
\end{align}
\end{subequations}
Here we used the following formula,
\begin{equation}
 \frac{1}{a\gamma^0 \Lambda_+ +b\gamma^0 \Lambda_-} =
  \frac{1}{a}\gamma^0 \Lambda_- +\frac{1}{b}\gamma^0 \Lambda_+\;,
\end{equation}
where $a$ and $b$ are arbitrary variables. By using these expressions, the spectral function $\rho$ is rewritten as follows:
\begin{equation}
 \rho(p) =
  \rho_{+}(p)\gamma^{0}\Lambda_{+}+\rho_{-}(p)\gamma^{0}\Lambda_{-}
  \label{eq:spectral_ft2}\;,
\end{equation}
where the coefficients $\rho_{\pm}(p)$ are defined by 
\begin{equation}
 \rho_{\pm}(p) \equiv \frac{-2 \; \text{\rm
  Im}\Sigma_{\mp}}{(p_+^{0}{\mp}p-\text{\rm
  Re}\Sigma_{\mp})^2+(\text{\rm Im}\Sigma_{\mp})^2}\
  \label{eq:spectral_ft3}\;.
\end{equation}

Thus the spectral function is written in terms of self energy. In other words, our calculation of the shear viscosity is reduced to that of the quark self energy. It is instructive to notice that if there is no interaction, the spectral function is simply given by 
\begin{equation}
 \rho^{(0)}({\bf p},\varepsilon) = 2\pi\delta(p_+^{0}-p)\gamma^{0}\Lambda_{+}+2\pi\delta(p_+^{0}+p)\gamma^{0}\Lambda_{-}
  \label{eq:viscosity_final2}.
\end{equation}
In this case, the viscosity $\eta$ obviously diverges, which indicates that the quark interaction is essential for realizing finite viscosity.

\section{\label{soft_modes}Soft Modes in the Quark Matter}

\begin{figure}
 \includegraphics[width=\linewidth]{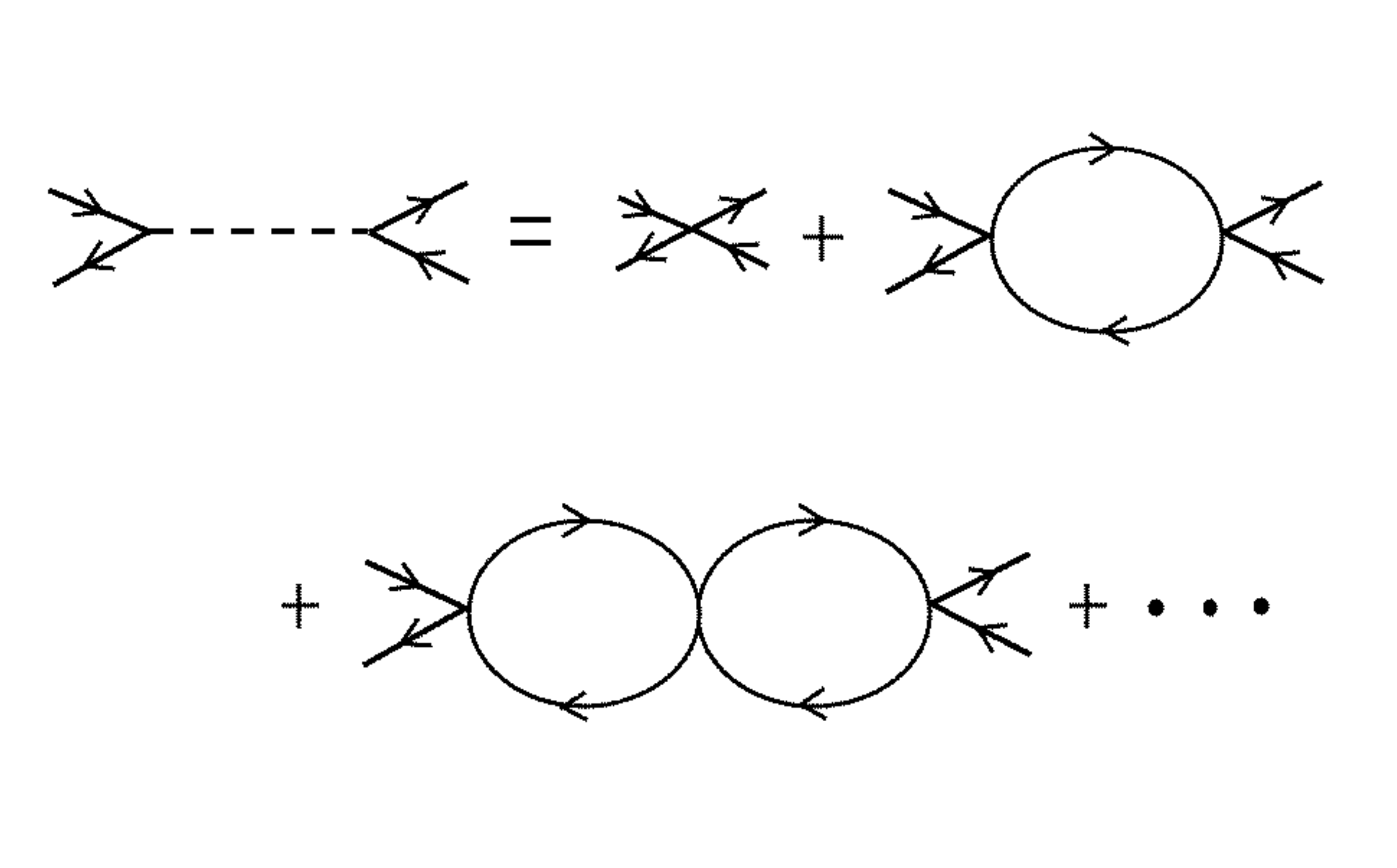}
  \caption{The soft modes.}
 \label{fig:soft_modes}
\end{figure}

In order to obtain finite shear viscosity, we have to know the self energy of quark $\Sigma$. As the largest contribution to the self energy, we assume that the self energy is acquired by the coupling to the soft modes, which are predicted in the scheme of the NJL model \cite{HK85},\cite{KKN05}. The soft mode is a collective mode of quark-antiquark pair above $T_c$ as shown in Fig.\ref{fig:soft_modes}. 
The existence of such mode is one of the precursor phenomena of the chiral transition and it behaves as a bosonic quasi-particle. As the soft modes are carriers of chirality, isospin and energy momentum, it is likely that the modes affect various transport coefficients including viscosity.

The propagator represented by Fig.\ref{fig:soft_modes} is written as
\begin{eqnarray}
D^{\alpha}(q)&=&D^{(0)}(1+X^{\alpha}D^{(0)}+X^{\alpha}D^{(0)}X^{\alpha}D^{(0)}+\cdot\cdot\cdot) \nonumber\\
&=& \frac{D^{(0)}}{1-D^{(0)}X^{\alpha}(q)}\;\label{eq:D},
\end{eqnarray}
with $D^{(0)}=-2g$. The factor $2$ in this equation is a symmetry factor
of the interaction $(\bar\psi\Gamma^{\alpha}\psi)^2$. The polarization operator $X^{\alpha}$ is defined by
\begin{align}
 X^{\alpha}&({\bf q},i\omega_n) \equiv T\sum_l \int \frac{\mathrm{d}^3 p}{(2\pi)^3}\nonumber \\
 &\times \text{\rm Tr}[\Gamma^{\alpha} G^{(0)}({\bf p+q},i\varepsilon_l+i\omega_n) \Gamma^{\alpha} G^{(0)}({\bf p},i\varepsilon_l)]\;
 \label{eq:X}.
\end{align}
Here $\Gamma^{0}=1$, $\Gamma^{\alpha}=i\gamma_{5} \tau^{\alpha} (\alpha=1,2,3)$ and the trace should be summed up over all Lorentz, color and flavor indices. Using the relation, $\Gamma^{\alpha}\gamma^{\mu}\Gamma^{\alpha}=\gamma^{\mu}$, the above polarization operator and the propagator of the soft mode are independent of $\alpha$: $X^{\alpha}(q)\equiv X(q)$ and $D^{\alpha}(q)\equiv D(q)$. The spectral representation of $G^{(0)}$ is read as
\begin{equation}
G^{(0)}({\bf k},i\varepsilon_m)=\int_{-\infty}^{\infty}\frac{\mathrm{d}\varepsilon}{2\pi}\frac{\rho^{(0)}({\bf k},i\varepsilon)}{i\varepsilon_m-\varepsilon}\;.
\end{equation}
Substituting this representation into Eq.(19), we obtain 
\begin{align}
 X({\bf q},i\omega_n) \equiv &T\sum_l \int \frac{\mathrm{d}^3
 p}{(2\pi)^3}\int\frac{\mathrm{d}\varepsilon}{2\pi}\int\frac{\mathrm{d}\varepsilon^{\prime}}{2\pi}\nonumber
 \\
 &\times \frac{1}{(i\varepsilon_l+i\omega_n-\varepsilon)(i\varepsilon_l-\varepsilon^{\prime})}\nonumber
\\
 &\times
\mathrm{Tr}[\rho^{(0)}({\bf p}+{\bf q},\varepsilon)\rho^{(0)}({\bf
 p},\varepsilon^{\prime})]\;.
\end{align}
Using the summation formula of Matsubara frequencies,
\begin{align}
 T\sum_l&
 \frac{1}{(i\varepsilon_l+i\omega_n-\varepsilon)(i\varepsilon_l-\varepsilon^{\prime})}\nonumber
 \\
 &=\frac{1}{2}\frac{1}{\varepsilon-\varepsilon^\prime-i\omega_n}\left(\tanh\frac{\varepsilon^{\prime}}{2T}-\tanh\frac{\varepsilon}{2T}\right)\;,
\end{align}
we obtain the expression of the polarization operator,
\begin{align}
 X({\bf q},i\omega_n) =&
 \int\frac{\mathrm{d}^{3}p}{(2\pi)^3}\int\frac{\mathrm{d}\varepsilon}{2\pi}\int\frac{\mathrm{d}\varepsilon^{\prime}}{2\pi}\frac{1}{\varepsilon-\varepsilon^{\prime}-i\omega_n}\nonumber
 \\
 &\times \frac{1}{2}\left(\tanh\frac{\varepsilon^{\prime}}{2T}-\tanh\frac{\varepsilon}{2T}\right)\nonumber
 \\
 &\times
 \mathrm{Tr}[\rho^{(0)}({\bf p}+{\bf q},\varepsilon)\rho^{(0)}({\bf p},\varepsilon^{\prime})]
 \label{eq:X_spectral}\;.
\end{align}

\begin{figure}
 \includegraphics[width=\linewidth]{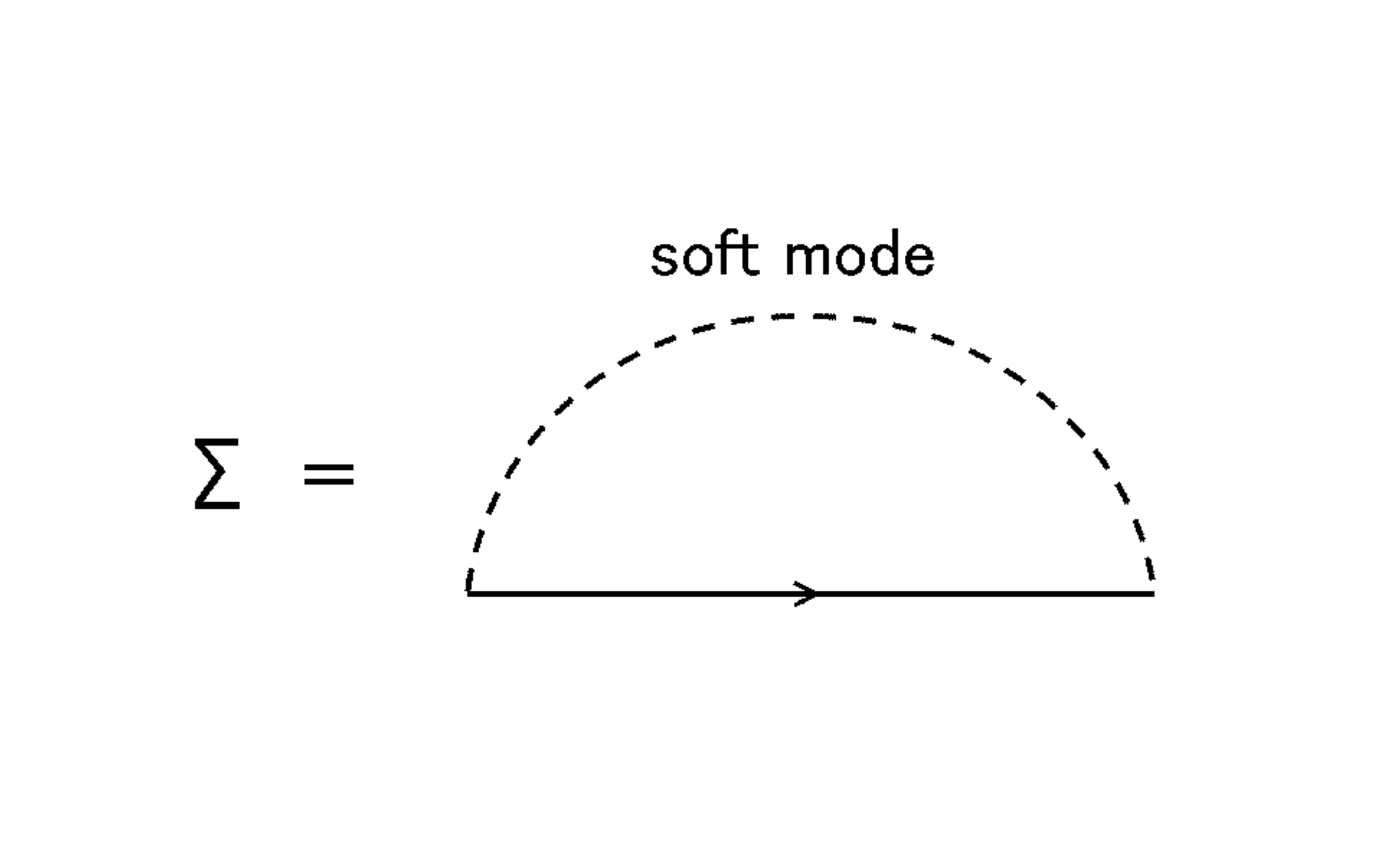}
  \caption{Soft mode coupling.}
 \label{fig:soft_mode_coupling}
\end{figure}

Now we consider the self energy of quark acquired by the coupling to the soft modes shown in Fig.\ref{fig:soft_mode_coupling}.
The self energy is written as
\begin{align}
 \Sigma({\bf p},i\varepsilon_l)=&\sum_{\alpha}T\sum_m\int\frac{\mathrm{d}^{3}k}{(2\pi)^3}\nonumber \\
 \times &\Gamma^{\alpha}G^{(0)}({\bf k},i\varepsilon_m)\Gamma^{\alpha}D({\bf p-k},i\varepsilon_l-i\varepsilon_m)\;,
\label{eq:sigma}
\end{align}
Here let us substitute the spectral representations of $G^{(0)}$ and $D$. The spectral representation of $G^{(0)}$ is given in Eq.(20) and that of $D$ is denoted by
\begin{equation}
 D({\bf q},i\omega_n)=\frac{1}{\pi}\int_{-\infty}^{\infty}\mathrm{d}\omega\frac{\mathrm{Im}D^{\text{R}}({\bf q},\omega)}{\omega-i\omega_n}\;
\label{eq:soft_mode_spectral_ft}.
\end{equation}
After these substitutions, the self energy Eq.(\ref{eq:sigma}) can be rewritten as, 
\begin{align}
 \label{eq:SE_rho} \Sigma({\bf p},i\varepsilon_l)
  =& \sum_{\alpha}T\sum_m\int\frac{\mathrm{d}^{3}k}{(2\pi)^3}\int\frac{\mathrm{d}\omega}{\pi}\int\frac{\mathrm{d}\varepsilon}{2\pi}\rho^{(0)}(k) \nonumber \\
   &\times \frac{\mathrm{Im}D^{\text{R}}(p-k)}{(i\varepsilon_m-\varepsilon)(\omega+i\varepsilon_m-i\varepsilon_l)}\;.
 \end{align}
Here again we made use of the relation $\Gamma^{\alpha}\gamma^{\mu}\Gamma^{\alpha}=\gamma^{\mu}$. Using another summation formula of Matsubara frequencies,
 \begin{align}
  T\sum_m & \frac{1}{(i\varepsilon_m-\varepsilon)(i\varepsilon_m+\omega-i\varepsilon_l)}\nonumber
  \\
  &= \frac{1}{2}\frac{1}{i\varepsilon_l-\omega-\varepsilon}\left(\coth\frac{\omega}{2T}+\tanh\frac{\varepsilon}{2T}\right)\;,
 \end{align}
we get the expression,
\begin{eqnarray}
\Sigma({\bf p},i\varepsilon_l) &=& -4\int\frac{{\mathrm{d}^3}k}{(2\pi)^3} \int\frac{{\mathrm{d}}\varepsilon}{2\pi} \int \frac{{\mathrm{d}}\omega}{2\pi}\left(\tanh\frac{\varepsilon}{2T}+\coth\frac{\omega}{2T}\right) \nonumber\\
& &\times\frac{\rho^{(0)}({\bf k},\varepsilon)\text{\rm Im}D^{\text{R}}({\bf p}-{\bf k},\omega)}{\omega+\varepsilon-i\varepsilon_l},
\end{eqnarray}
where the factor $4$ comes from the number of the soft modes.
 Here it is noted that the principal value for $\coth(\omega/2T)$ should be taken in the integral. If we substitute the spectral function Eq.(17) given in the previous section into Eq.(\ref{eq:SE_rho}), the self energy is read as
\begin{eqnarray}
  \Sigma({\bf p},i\varepsilon_l) = -4\int\frac{\mathrm{d}^{3}k}{(2\pi)^3}\int\frac{{\mathrm{d}}\omega}{2\pi}\frac{\text{\rm Im}D^{\text{R}}({\bf p}-{\bf k},\omega)}{2k} \nonumber \\
 \times \bigg[\left(\coth\frac{\omega}{2T}+\tanh\frac{k-\mu}{2T}\right)\frac{k_+ \cdot \gamma}{\omega+k-\mu-i\varepsilon_l} \nonumber \\
 +\left(\coth\frac{\omega}{2T}-\tanh\frac{k+\mu}{2T}\right)\frac{k_-\cdot \gamma}{\omega-k-\mu-i\varepsilon_l}\bigg]\;,
\end{eqnarray}
where we have introduced four-vector $k_{\pm} \equiv ({\bf k},\pm k)$. If the $z$-axis is taken in parallel with ${\bf p}$, the integral over $k_x$ $(k_y)$ vanishes $(k_{\pm}\cdot\gamma \rightarrow \pm k\gamma^{0}-k_z \gamma^3)$ and yields
\begin{eqnarray}
 \Sigma({\bf p},i\varepsilon_l) = -4\int\frac{\mathrm{d}^{3}k}{(2\pi)^3}\int\frac{{\mathrm{d}}\omega}{2\pi}\frac{\text{\rm Im}D^{\text{R}} ({\bf p}-{\bf k},\omega)}{2} \nonumber \\
  \times \bigg[\left(\coth\frac{\omega}{2T}+\tanh\frac{k-\mu}{2T}\right)\frac{\gamma^{0}-(\hat{\bf p}\cdot\hat{\bf k})\hat{\bf p}\cdot\bm{\gamma}}{\omega+k-\mu-i\varepsilon_l} \nonumber \\
  +\left(\coth\frac{\omega}{2T}-\tanh\frac{k+\mu}{2T}\right)\frac{\gamma^{0}+(\hat{\bf p}\cdot\hat{\bf k})\hat{\bf p}\cdot\bm{\gamma}}{\omega-k-\mu-i\varepsilon_l}\bigg].
\end{eqnarray}
This equation is a linear combination of the following form:
\begin{equation}
 \Sigma({\bf p},i\varepsilon_l)=\Sigma_{0}(p)\gamma^{0}+\Sigma_{1}(p)\hat{\bf p}\cdot\bm{\gamma}.
\end{equation}
The first and second terms represent the scalar and vector parts of the self energy respectively.

\section{\label{approximation}Approximation Method}

Before calculating the shear viscosity, it is appropriate to carry out the following approximation without loss of the physical essence. In our calculations we have three-dimensional integration containing $\lvert {\bf p}\pm{\bf q} \rvert$. Then we take the following approximation:
\begin{equation}
 \lvert {\bf p}\pm{\bf q} \rvert \simeq
 p\theta(p-q)+q\theta(q-p)\nonumber \equiv \kappa(p,q)\;,
\end{equation}
where $\theta$ denotes usual step function: $\theta(x)=1\, (x>0)$ and $\theta(x)=0\, (x<0)$. This approximation is to replace the scalar product ${\bf p}\cdot{\bf q}$ by its average. From the above approximation, ${\bf p}\cdot{\bf q}$ vanishes because ${\bf p}\cdot{\bf q}=[({\bf p}+{\bf q})^{2}-({\bf p}-{\bf q})^{2}]/4 \simeq 0$. With the use of this approximation, we can proceed our calculations furthermore.

\subsection{Soft mode}

The polarization operator $X({\bf q})$ in Eq.(\ref{eq:X_spectral}) is represented by multiple integral. Using our approximation, the trace of Eq.({\ref{eq:X_spectral}}) can be written as follows:
\begin{align}
 \text{\rm Tr}&[\rho^{(0)}({\bf p}+{\bf q},\varepsilon)\rho^{(0)}({\bf p},\varepsilon^{\prime})] \simeq \frac{4\pi^2}{p\kappa(p,q)}[(\varepsilon+\mu)(\varepsilon^{\prime}+\mu)-p^2]  \nonumber \\
 &\times [\delta(\varepsilon+\mu-\kappa(p,q))-\delta(\varepsilon+\mu+\kappa(p,q))] \nonumber \\
 &\times [\delta(\varepsilon^{\prime}+\mu-p)-\delta(\varepsilon^{\prime}+\mu+p)]\;.
\end{align}
Carrying out the angular integration of Eq.(\ref{eq:X_spectral}), we get
\begin{equation}
 X(q,i\omega_m) = \frac{N_{c}N_{f}}{2\pi^2}\int_o^{\infty} S(p,q)(n_{p}+m_{p}-1)p^{2}{\mathrm d}p,
\label{eq:X_S}
\end{equation}
where $S(p,q)$ is defined by
\begin{eqnarray}
  S(p,q) &\equiv& \frac{q-p}{q}\left(\frac{1}{p-q-i\omega_m}+\frac{1}{p-q+i\omega_m}\right)\theta(q-p) \nonumber \\
 &+& \frac{q+p}{q}\left(\frac{1}{p+q-i\omega_m}+\frac{1}{p+q+i\omega_m}\right)\theta(q-p) \nonumber \\
 &+& \left(\frac{2}{2p-i\omega_m}+\frac{2}{2p+i\omega_m}\right)\theta(p-q).
\end{eqnarray}
In the right hand side, the $n_p \equiv [e^{\beta(p-\mu)}+1]^{-1}$ and $m_p \equiv [e^{\beta(p+\mu)}+1]^{-1}$ denote the quark and antiquark distribution functions, respectively. Thus we can calculate the real part $(\text{\rm Re}X^R)$ and the imaginary part $(\text{\rm Im}X^R)$ of $X^R$ from this equation by using the relation, $1/(x+i\delta)=P/x-i\pi\delta(x)$. From Eq.({\ref{eq:D}}), the imaginary part of $D^{\text{R}}$ is read as
\begin{equation}
 \text{\rm Im}D^{\text{R}}(q,\omega) = \frac{-4g^2\text{\rm Im}X^R(q)}{(1+2g\text{\rm Re}X^R(q))^{2}+4g^{2}(\text{\rm Im}X^R(q))^2}\
\label{eq:Im_D}.
\end{equation}
This equation plays a fundamental role in the later discussion as far as the soft mode is concerned.

\subsection{Self energy of quark}

Next we go on to the calculation of the self energy $\Sigma(p)$. From the above equation, we substitute ${\bf p}\cdot{\bf k}=0$ into Eq.(30), the vector part of the self energy $\Sigma_{1}$ vanishes. Hence we get
\begin{equation}
 \Sigma \simeq \Sigma_{0}(p)\gamma^0 \;.
\end{equation}
This means that the self energy is independent of the direction of the momentum. The final expression for $\Sigma_{0}(p)$ is
\begin{eqnarray}
 \label{eq:Sigma_0}
 \Sigma_{0}(p,i\varepsilon_l) \equiv -2\int\frac{\mathrm{d}^{3} k}{(2\pi)^3}\int\frac{\mathrm{d}\omega}{2\pi}\text{\rm Im}D^{\text{R}}(\kappa(p,k),\varepsilon)\nonumber \\
 \times \bigg[\left(\coth\frac{\omega}{2T}+\tanh\frac{k-\mu}{2T}\right)\frac{1}{\omega+k-\mu-i\varepsilon_l}\nonumber \\
 +\left(\coth\frac{\omega}{2T}-\tanh\frac{k+\mu}{2T}\right)\frac{1}{\omega-k-\mu-i\varepsilon_l}\bigg]\;.
\end{eqnarray}
The retarded function $\Sigma_{0}^{\text{R}}$ is obtained through analytic continuation $(i\varepsilon_l \rightarrow \varepsilon+i\delta)$ of this equation. Its imaginary part is given as follows:
\begin{align}
 &\text{\rm Im}\Sigma_{0}^{\text{R}}(p,\varepsilon)  = 
  \frac{-1}{2\pi^2}\int_{0}^{\infty}\text{\rm d}k k^{2}\nonumber
 \\
 &\times
 \bigg[\left(\coth\frac{\omega-k+\mu}{2T}+\tanh\frac{k-\mu}{2T}\right)\text{\rm
 Im}D^{\text{R}}(\kappa,\omega-k+\mu)\nonumber
 \\
 &+\left(\coth\frac{\omega+k+\mu}{2T}-\tanh\frac{k+\mu}{2T}\right)\text{\rm Im}D^{\text{R}}(\kappa,\omega+k+\mu)\bigg]\;.
\end{align}

As for the calculation of the real part $\text{\rm Re}\Sigma_{0}^{\text{R}}$, which is the potential energy of quark, we make use of the dispersion relation (Hilbert transformation) as follows:
\begin{equation}
 \text{\rm Re}\Sigma_{0}^{\text{R}}(p,\varepsilon) = \frac{1}{\pi}\int_{-\infty}^\infty\frac{\text{\rm Im}\Sigma_{0}^{\text{R}}(p,\varepsilon^{\prime})}{\varepsilon^{\prime}-\varepsilon}\mathrm{d}\varepsilon^{\prime},
\end{equation}
where we should take the principal value of the integral in the right hand side. Now that we have obtained the self energy $\Sigma$, we can calculate the shear viscosity using Eq.(7).

\section{\label{results}NUMERICAL RESULTS}

We are now ready to evaluate the shear viscosity at various temperature and density. The parameters of the NJL model are taken to be the same values as those in the reference \cite{HK94}: the coupling constant, $g=5.2\, {\rm MeV}$ and the cut off parameter, $\Lambda=631\,{\rm MeV}$. The quark mass $m$ is assumed to be zero so that the chiral symmetry holds exactly. First we restrict ourselves to the case of the chemical potential $\mu=0$. Note that the RHIC data covers $\mu\sim 10\,{\rm MeV}$. Under these parameters, the critical temperature $T_{c}$ is $161\,{\rm MeV}$, which is derived from the vanishing condition of the inverse propagator of the soft mode: $D^{-1}(p=0)=0$ \cite{HK94}.

\subsection{Soft mode}

To begin with, let us consider the soft mode in the quark matter. Its spectral functions ${\rm Im}D^{\text{R}}(p=0,\omega)$ are calculated from Eq.(35) and the results are shown in Fig.4 at various temperatures: $T=180,\: 210,\: 260\,{\rm MeV}$. For $T$ very close to $T_{c}$, the soft mode becomes a sharp resonance and acquires the nature of a good quasi-particle. As the temperature increases, the resonance width is spread out and its strength is weakened at $170\,{\rm MeV}< T <260\,{\rm MeV}$.
\begin{figure}
 \includegraphics[width=\linewidth]{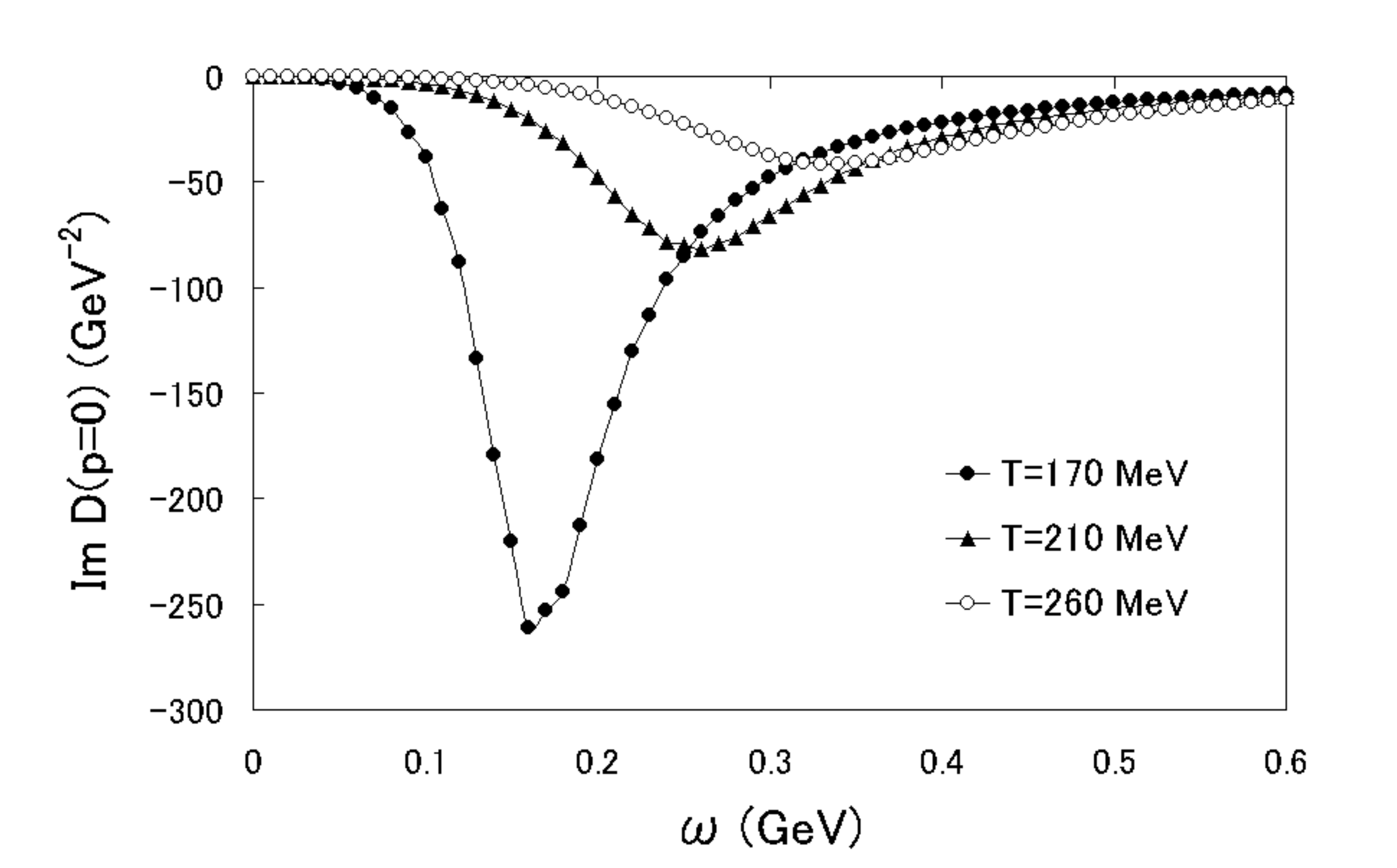}
 \caption{The spectral function of the soft mode at $T=170, 210, 260\,{\rm MeV}$ as a function of the frequency of the soft mode.}
\end{figure}
As is pointed out by Hatsuda and Kunihiro, this peak near the critical temperature reflects a precursor of the chiral phase transition. Therefore its existence implies that the system is in the restored chiral symmetry phase or QGP phase. In Fig.5, we present the spectral functions with various momenta at $T=170\,{\rm MeV}$. 
\begin{figure}
 \includegraphics[width=\linewidth]{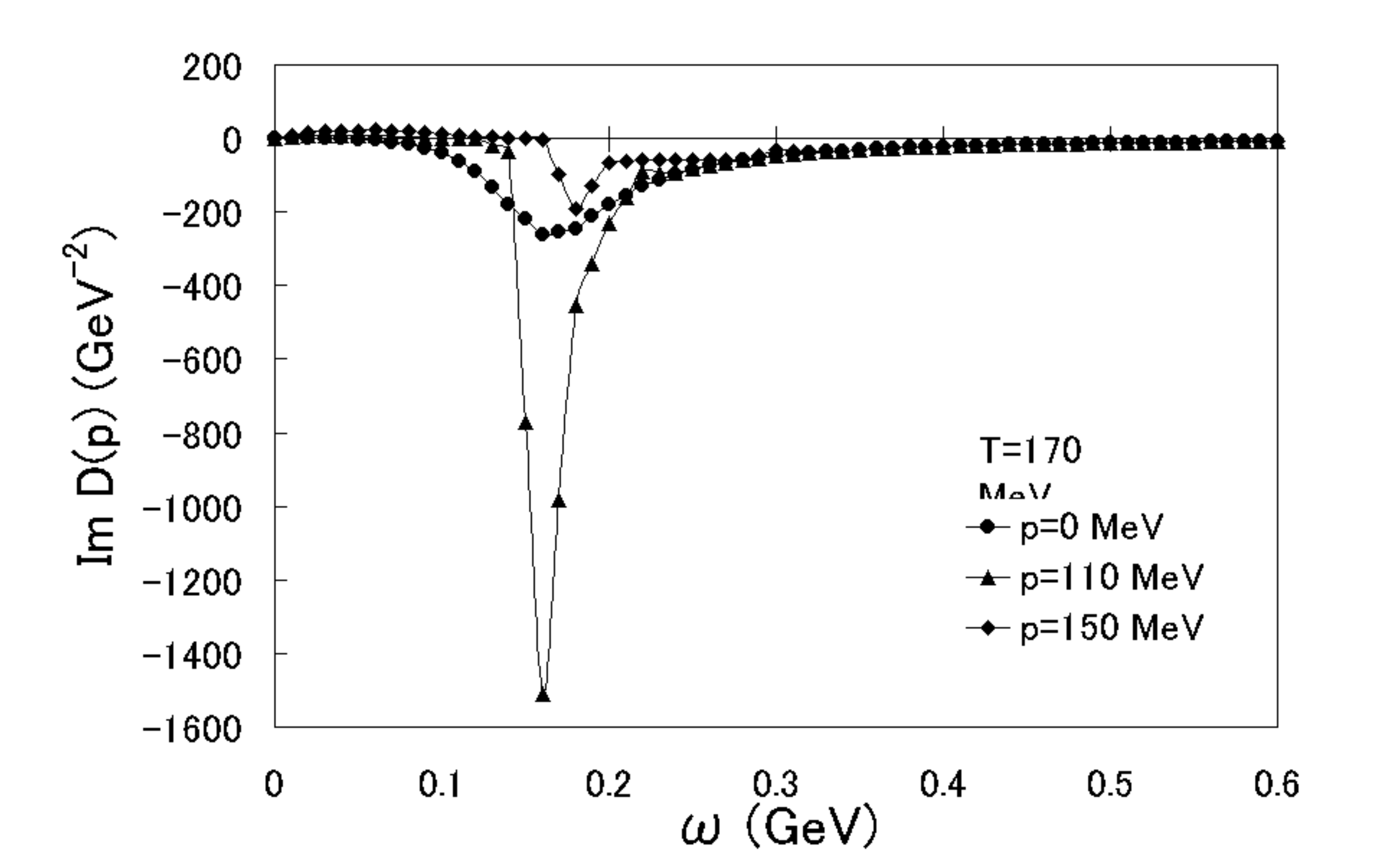}
 \caption{The spectral function of the soft mode at the momenta of quark, $p=0, 110, 150\,{\rm MeV}$ as a function of the frequency of the soft mode ($T=170\,{\rm MeV}$).}
\end{figure}
The position of the peak does not depend on the momentum $p$ but the strength depends on $p$ or the temperature $T$. Hence the soft mode is determined by the temperature. It will be shown later that the soft mode is not sensitive to the chemical potential.

\subsection{Self energy of quark}

Next let us calculate the self energy of the quark, which is described by the coupling to the soft modes as discussed in the previous section. The real and imaginary parts of $\Sigma$ do not depend on the momentum $p$ sensitively as well as the soft mode. We show the spectral function of quark in Fig.6.
\begin{figure}
 \includegraphics[width=\linewidth]{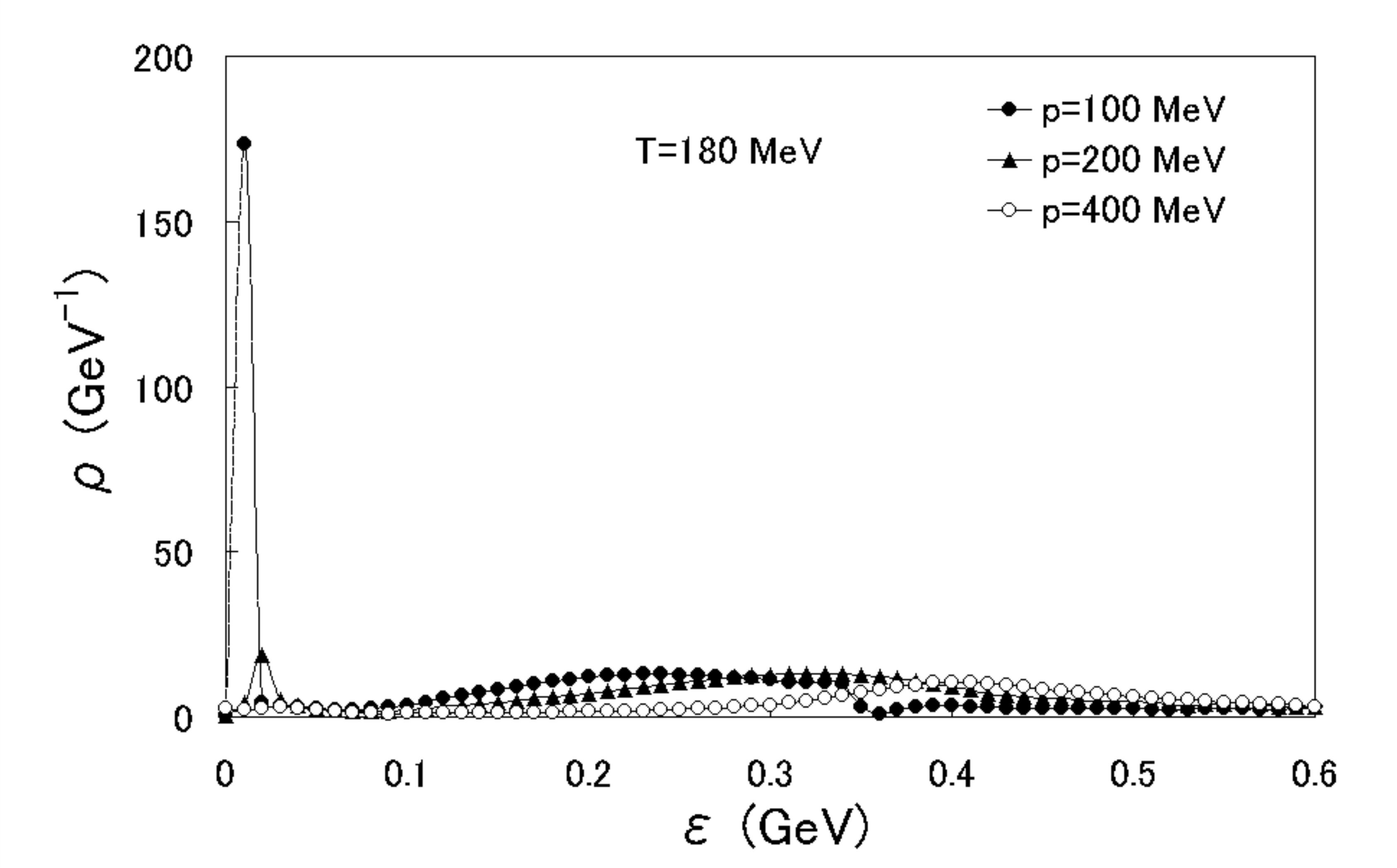}
 \caption{The spectral function of the quark at the momenta, $p=0, 200, 400\,{\rm MeV}$ ($T=180\,{\rm MeV}$), which is generated by the coupling to the soft mode.}
\end{figure}
They have a broad and small peak except for the case of $p=100\,{\rm MeV}$. The position of the broad peak moves with the momentum such that the denominator in Eq.(16) is minimized. The sharp peak near $\varepsilon\sim 0$ at $p=100\,{\rm MeV}$ seems to come from the overlap between two spectral functions, $\rho^{(0)}$ and ${\rm Im}D^{\text{R}}$. We see that the strength of the spectral functions are suppressed due to the integration of the momentum $k$. This behavior lowers the shear viscosity as shown in the next subsection.

\subsection{Shear viscosity}

Now we can calculate the shear viscosity by using Eq.(7). The results are shown in Fig.7 as a function of the temperature above the critical one, $T>T_{c}=161{\rm MeV}$. Similarly its ratio to the entropy is drawn in Fig.8. The solid lines in these figures represent guidelines for the data.
\begin{figure}
 \includegraphics[width=\linewidth]{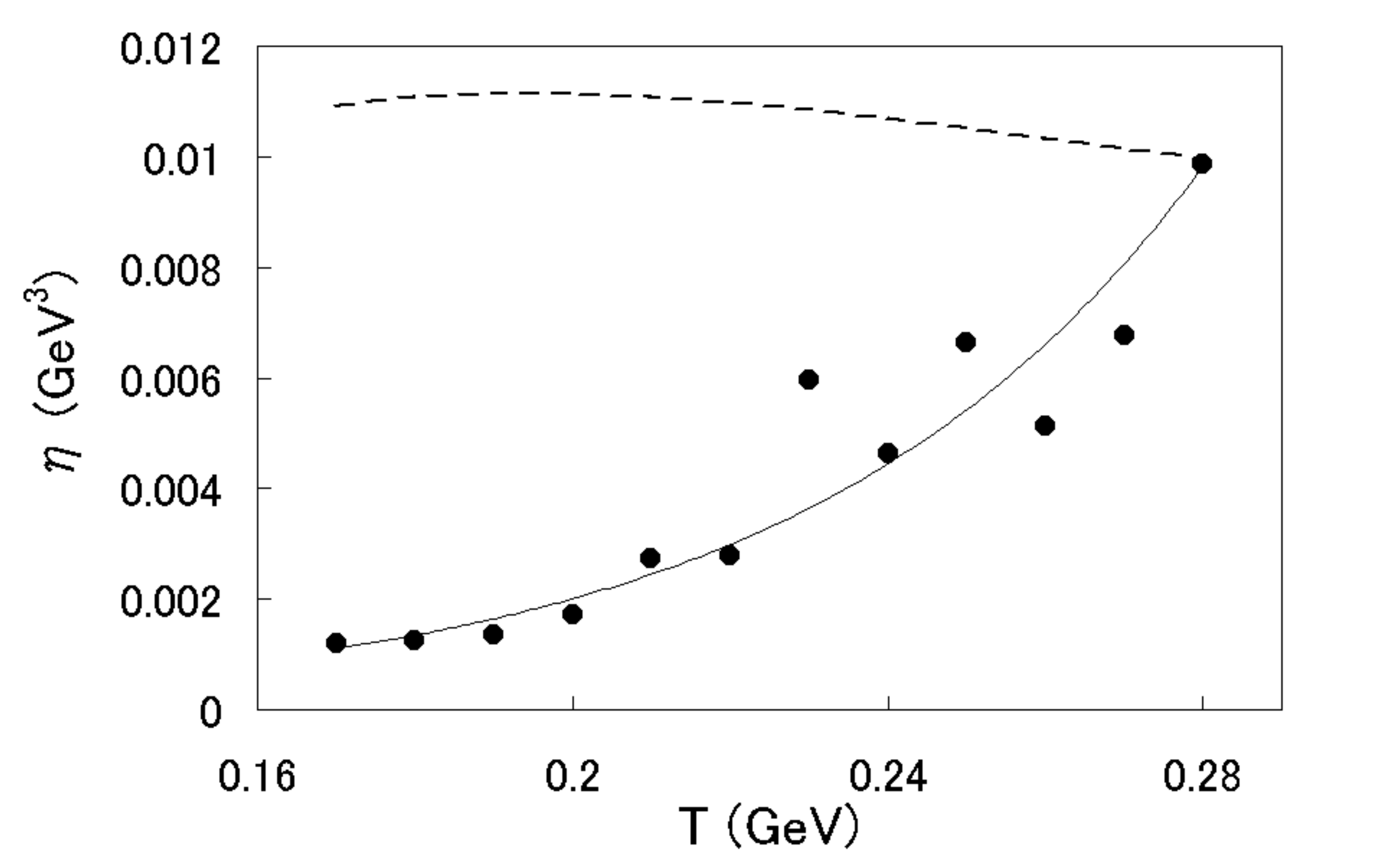}
 \caption{The shear viscosity as a function of the temperature $T$ ($\mu=0$). The dashed line denotes for result by the particle mode explained in the text.}
\end{figure}
\begin{figure}
 \includegraphics[width=\linewidth]{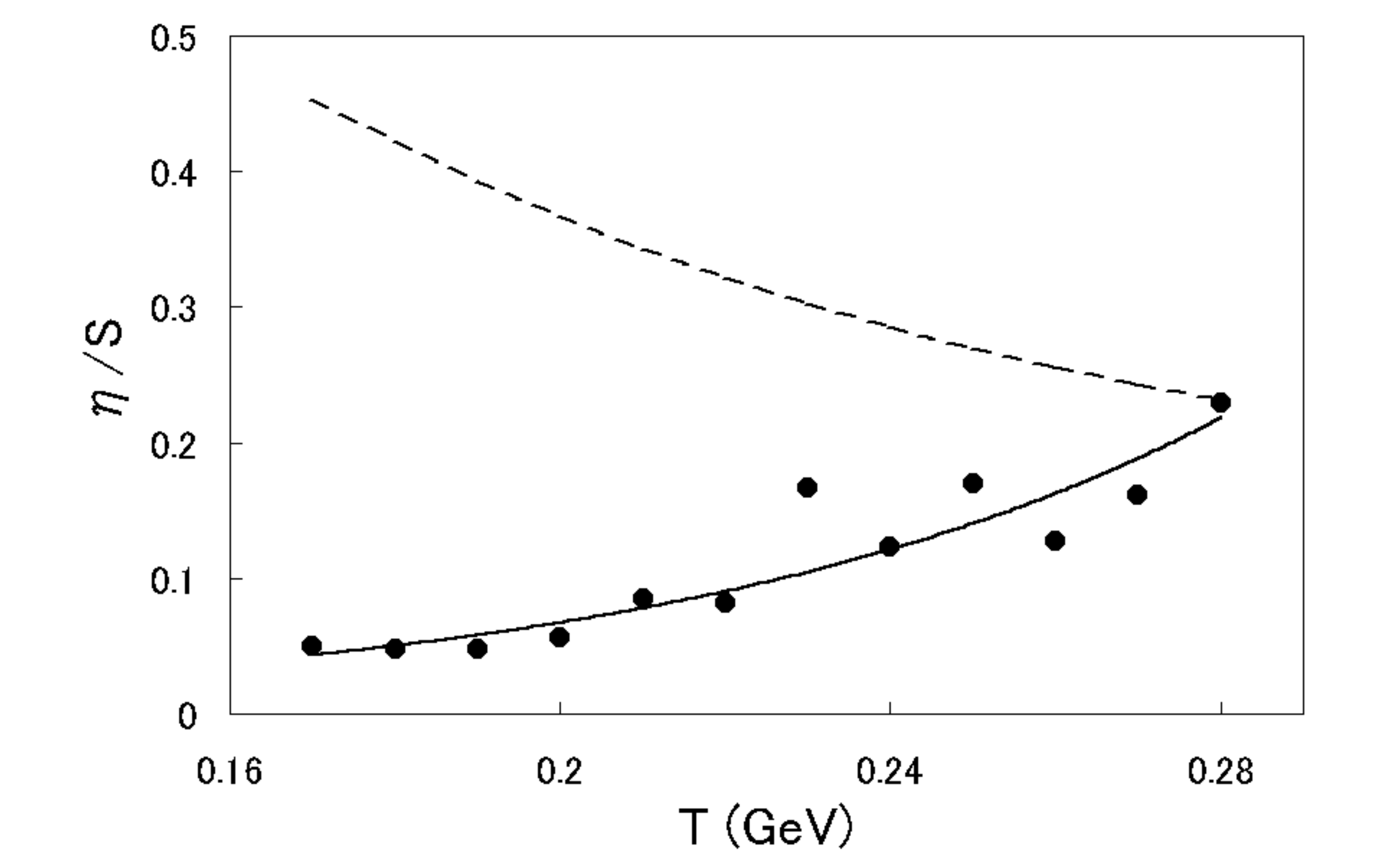}
 \caption{The ratio of the shear viscosity to the entropy as a function of the temperature $T$ ($\mu=0$). The dashed line denotes for result by the particle mode explained in the text.}
\end{figure}
These values are almost constant but increase gradually at $T>200\,{\rm MeV}$. The smallness of the shear viscosity is understood by the smallness of the quark spectral function as shown in Fig.6. This comes from the strong coupling to the soft modes, which gives rise to the large self energy, that is the short mean free path. As the temperature increases, the soft mode becomes weak and its peak moves to the high frequency (Fig.4). Then the quark spectral function is enhanced. This is the reason that the shear viscosity increases gradually at $T=280\,{\rm MeV}$.

For the comparison, we also calculated the shear viscosity by the coupling to the particle mode. This mode represents that the quark collides with another particle instead of the soft mode. The Feynman diagram of this process corresponds to the second diagram of the Fig.3. Hence the spectral function of this mode is given by
\begin{equation}
{\rm Im}D^R(q)=4(2g)^{2}{\rm Im}X^R(q).
\end{equation}
The calculated result is denoted by the dashed line in Fig.7 and 8. We see that the viscosity by the particle mode is one-order larger than those by the soft mode. This means that the particle mode gives rise to the weak correlation between quarks and makes the quark matter Fermi gas-like. This property is seen in the spectral function of the particle mode as shown in Fig.9.
\begin{figure}
 \includegraphics[width=\linewidth]{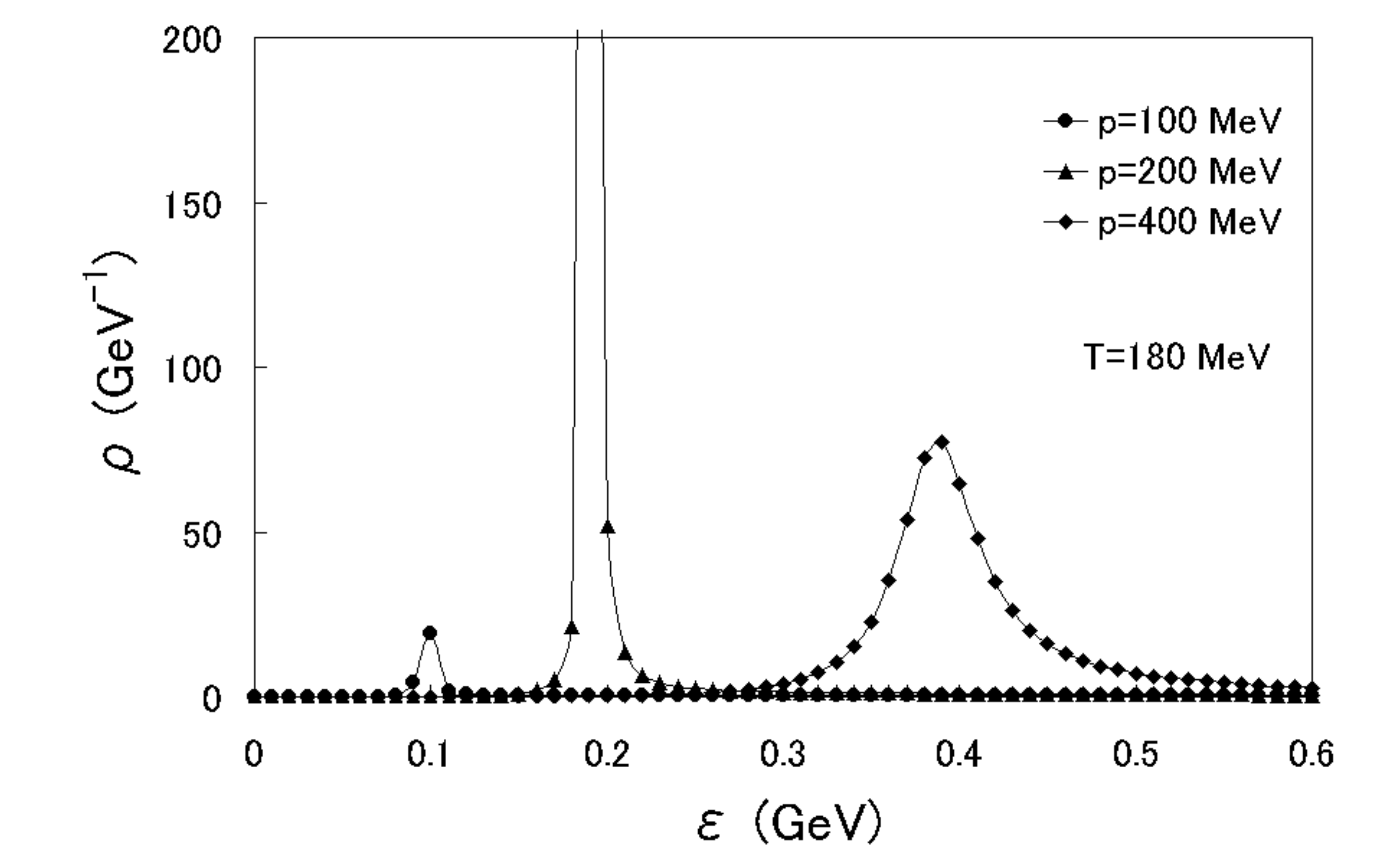}
 \caption{The spectral function of the quark at the momenta, $p=0, 200, 400\,{\rm MeV}$ ($T=180\,{\rm MeV}$), which is generated by the coupling to the particle mode.}
\end{figure}

\subsection{Finite density}

Next let us discuss the density or chemical potential dependence of the shear viscosity. The shear viscosity for $0<\mu<100\,{\rm MeV}$ is shown in Fig.10. The viscosity is increasing very slowly with the chemical potential. This dependence is understood by noting that the soft mode is not changed for the chemical potential as shown in Fig.11 at $T=170\,{\rm MeV}$.
\begin{figure}
 \includegraphics[width=\linewidth]{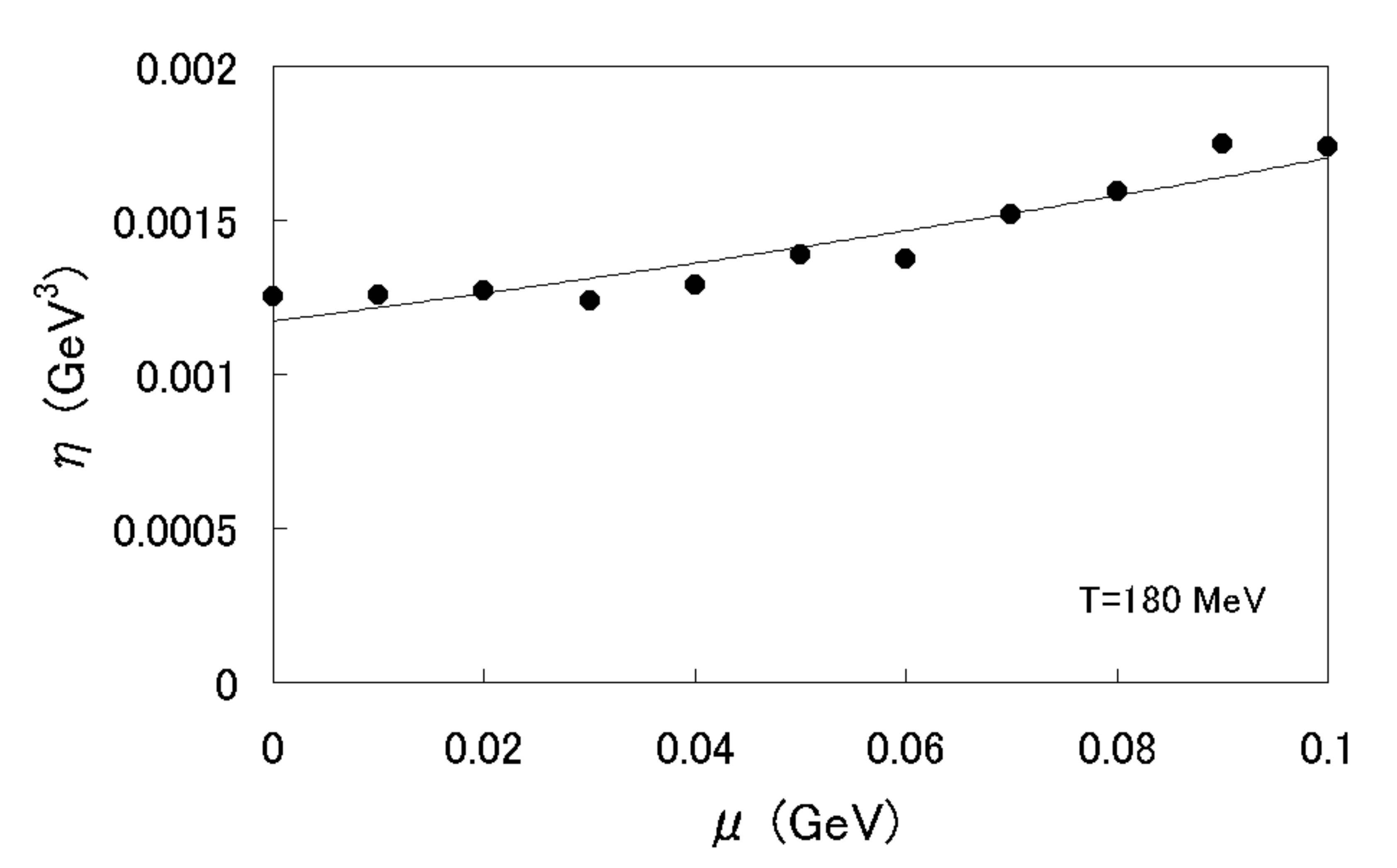}
 \caption{The shear viscosity as a function of the chemical potential $\mu$ ($T=180\,{\rm MeV}$).}
\end{figure}
\begin{figure}
 \includegraphics[width=\linewidth]{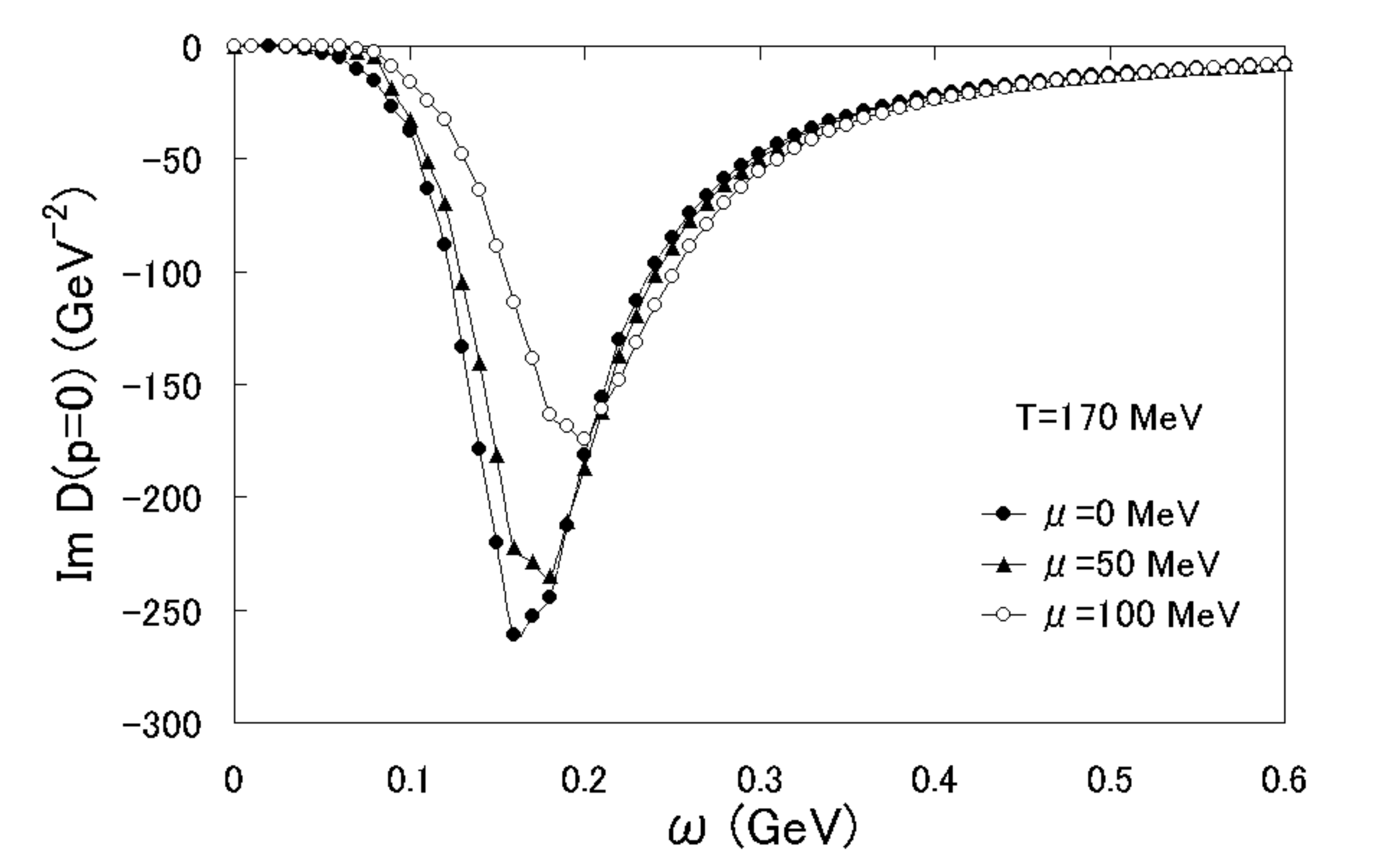}
 \caption{The spectral function of the soft mode at the chemical potential, $\mu=0, 50, 100\,{\rm MeV}$ as a function of the frequency of the soft mode ($T=170\,{\rm MeV}$).}
\end{figure}

\subsection{Reynolds number}

Finally we will discuss how small the shear viscosity obtained in this paper is. To this end, let us introduce (dimensionless) Reynolds number which is defined by
\begin{equation}
R=\frac{{\rm inertia\,term}}{{\rm viscosity\,term}}=\frac{l\cdot u}{\eta/(\rho_{m})}.
\end{equation}
In the numerator, $l$ and $u$ denote typical size and velocity of the system, respectively. We set $l=10\,{\rm fm}$ and $u=1={\rm light\: velocity}$ for convenience. It is difficult to define ``mass density $\rho_{m}$" appeared in the denominator. Here we have defined it by the qusi-particle mass $m$ times number density of the quark and antiquark. If we set $m=10\sim 100\,{\rm MeV}$, the Reynolds number becomes
\begin{equation}
R=3\sim 30.\quad   ({\rm for\: soft\: mode})
\end{equation}
On the other hand, $R=0.3\sim 3$ in the case of the coupling to the particle mode. Taking into account that the shear viscosity is negligible under the large $R$, the evaluated $R$ implies that our quark matter is close to a perfect liquid due to the strong coupling to the soft mode.

\subsection{Summary}

In conclusion, we have evaluated the shear viscosity of the quark matter (QGP) by using the Kubo formula and the NJL model. The self energy of quark is calculated by the coupling to the soft mode which is a collective mode composed of the quark-antiquark pair. This strong correlation of quarks lowers the shear viscosity of the quark matter and the quark matter is nearly a perfect fluid. As the temperature increases, the shear viscosity increases gradually because the strength of the soft modes decreases with the temperature. On the other hand it is shown that the shear viscosity is not sensitively depend on the chemical potential.

\section*{Acknowledgments}

The authors would like to thank Professors Y.Tsue and K.Iida (Kochi University) for valuable comments and discussions. They also thank the Yukawa Institute for Theoretical Physics at Kyoto University. Discussions during the YITP workshop YITP-W-04-07 on Thermal Quantum Field Theories and Their Applications and the YKIS2006 on "New Frontiers QCD" were very useful to start and complete this work.

\end{document}